\documentclass{aastex}          
\usepackage{spr-astr-addons}    
\usepackage{epsfig}
\usepackage{graphicx}
\usepackage{color}

\begin{document}

\title{The study of magnetic reconnection in solar spicules}

\author{Z.~Fazel, and H.~Ebadi\altaffilmark{1}}
\affil{Astrophysics Department, Physics Faculty,
University of Tabriz, Tabriz, Iran\\
e-mail: \textcolor{blue}{z$_{-}$fazel@tabrizu.ac.ir}}

\altaffiltext{1}{Research Institute for Astronomy and Astrophysics of Maragha,
Maragha 55134-441, Iran.}

\begin{abstract}

This work is devoted to study the magnetic reconnection instability under solar spicule conditions. Numerical study of the
resistive tearing instability in a current sheet is presented by considering the magnetohydrodynamic (MHD) framework.
To investigate the effect of this instability in a stratified atmosphere of solar spicules, we solve linear and non-ideal MHD equations in the $x-z$ plane.
In the linear analysis it is assumed that resistivity is only important within the current sheet, and the exponential growth of energies takes place faster as plasma resistivity increases. We are interested to see the occurrence of magnetic reconnection during the lifetime of a typical solar spicule.
\end{abstract}


\section{Introduction}
\label{sec:intro}
The mechanism of coronal heating is one of the major problems in solar physics. The magnetic structure of the corona
can play an important role on the problem of heating, so it should be necessary to study the converting of the magnetic energy to heat.
\\Magnetic reconnection plays a critical role in many astrophysical processes, e.g. particle acceleration and solar flares. Magnetic reconnection
is a topological change in the field which violates the frozen-flux condition of ideal magnetohydrodynamics (MHD). If a magnetic field can
leak across the plasma it can reach a lower energy state, in the case of a current sheet, it can undergo "tearing" into magnetic islands.
Tearing instability is responsible for many cases of magnetic field line reconnection process which converts magnetic energy into the thermal and
kinetic energy of plasma flows in astrophysical and laboratory plasmas \citep{Pri2000, Birn2007, Zweibel2009}. Nowadays emphasis in the tearing mode
instability is focused on physical effects that result in a much faster rate of field line reconnection.
\\\citet{Yokoyama1995} modeled X-ray and EUV jets and surges observed with $H\alpha$ in the chromosphere by performing a resistive $2D$ MHD simulation
of the magnetic reconnection occurring in the current sheet between emerging magnetic flux and overlying pre-existing coronal magnetic fields.
\emph{Hinode} observation revealed that jets are ubiquitous in the chromosphere \citep{Kosugi2007}.
\citet{De2007}, from \emph{Hinode} data estimated the energy flux carried by transversal oscillations generated by spicules. They indicated that
the calculated energy flux is enough to heat the quiet corona. \citet{He2009} based on \emph{Hinode/SOT} observations found the signatures of small scale
reconnection in spicules. They concluded that magnetic reconnection can exit kink waves along spicules.
\\\citet{Sweet1958, Parker1963, Petschek1964, Soward1982} introduced magnetic reconnection as the central process allowing for efficient magnetic to kinetic
energy conversion in solar flares and for interaction between the magnetized interplanetary medium and magnetosphere of Earth.
\\\citet{Tak2001} have investigated photospheric magnetic reconnection induced by convective
intensification of solar surface magnetic fields, performing 2.5-dimensional MHD numerical simulations. They concluded that in models of solar spicules,
upward propagating slow waves or Alfv\'{e}n waves are usually assumed as initial perturbations. The wave energies due to the reconnections are comparable
to those assumed in the spicule models. The photospheric magnetic reconnection might be one of the important causes of solar spicules.
\\Spicules, thin and elongated structures are one of the most pronounced features of the chromosphere. They are seen in spectral lines
at the solar limb at speeds of about $20-25$~km~s$^{-1}$ propagating from the photosphere into the magnetized low atmosphere of the sun \citep{Tem2009}.
Their diameter and length varies from spicule to spicule having the values from $400$~km to $1500$~km and from $5000$~km to $9000$~km, respectively.
Their typical lifetime is $5-15$ min, however some spicules may live for longer or shorter periods \citep{Tem2010}. The typical electron density at heights where the spicules are observed
is approximately $3.5\times10^{16}-2\times10^{17}$ m$^{-3}$, and their temperatures are estimated as $5000-8000$ K \citep{bec68, ster2000}.
\\In this paper we present numerical simulation of magnetic reconnection due to the tearing mode instability.
Section $2$ gives the basic equations and theoretical model. In section $3$ numerical results are presented and discussed, and a brief summary is
followed in section $4$.

\section{Theoretical modeling}
\subsection{Equilibrium}
\label{sec:theory}
The formal description of reconnection requires the choice of a dynamical model. We confine the discussion to magnetohydrodynamics
for a finite resistivity (resistive MHD). The corresponding basic equations for the non-ideal MHD in the plasma dynamics are:
\begin{equation}
\label{eq:mass} \frac{\partial \mathbf{\rho}}{\partial t} + \nabla
\cdot (\rho \mathbf{v}) = 0,
\end{equation}
\begin{equation}
\label{eq:momentum} \rho\frac{\partial \mathbf{v}}{\partial t}+
\rho(\mathbf{v} \cdot \nabla)\mathbf{v} = -\nabla p + \rho
\mathbf{g}+ \frac{1}{\mu_{0}}(\nabla \times \mathbf{B})\times
\mathbf{B}
\end{equation}
\begin{equation}
\label{eq:induction} \frac{\partial \mathbf{B}}{\partial t} = \nabla
\times(\mathbf{v} \times \mathbf{B})+ \eta\nabla^2\mathbf{B},
\end{equation}
\begin{equation}
\label{eq:divergence} \nabla \cdot \mathbf{B} = 0,
\end{equation}
\begin{equation}
\label{eq:state} p = \frac{\rho RT}{\mu}.
\end{equation}
where $\rho$ is mass density, $\mathbf{v}$ is flow velocity, $\mathbf{B}$ is the magnetic field, $p$ is the gas pressure, $R$ is the universal gas constant,
$\eta$ is constant resistivity coefficient, $\mu_{0}$ is the vacuum permeability, $\mu$ is the mean molecular weight.
We assume that spicules are highly dynamic with speeds that are significant fractions of the Alfv\'{e}n speed ($V_{A0} = B_{0}/\sqrt{\mu_{0}\rho_{0}}$).
\\A planar slab of a uniform plasma is embedded in a sheared force-free magnetic field and surrounded by two perfectly conducting boundaries
at $x$= $0, L_{x}$, where $L_{x}$ is the box size in the $x$ direction:
\begin{eqnarray}
\label{eq:perv}
\textbf{B}_{0} &=& B_{0y}(x)\hat{j} + B_{0z}(x)\hat{k}
\end{eqnarray}
with
\begin{eqnarray}
\label{eq:shear field}
 B_{0y}(x) &=& 0, \nonumber\\
 B_{0z}(x) &=& \tanh(\frac{x-z_{t}}{z_{w}})
\end{eqnarray}
where $z_{t}$=$L_{x}/2$ is the position of the middle of a spicule and $z_{w}$ is the thickness of the initial current sheet (the shear length of magnetic field
configuration). Since the equilibrium magnetic field is force-free, the pressure gradient is balanced by the gravity force, which is assumed to be $\textbf{g}$=$-g\hat{k}$ via this equation:
\begin{equation}
\label{eq:balance}
 -\nabla p_{0}(z) + \rho_{0}(z)\textbf{g}=0.
\end{equation}

\subsection{Perturbations}
\begin{figure}
\centering
\includegraphics[width=8cm]{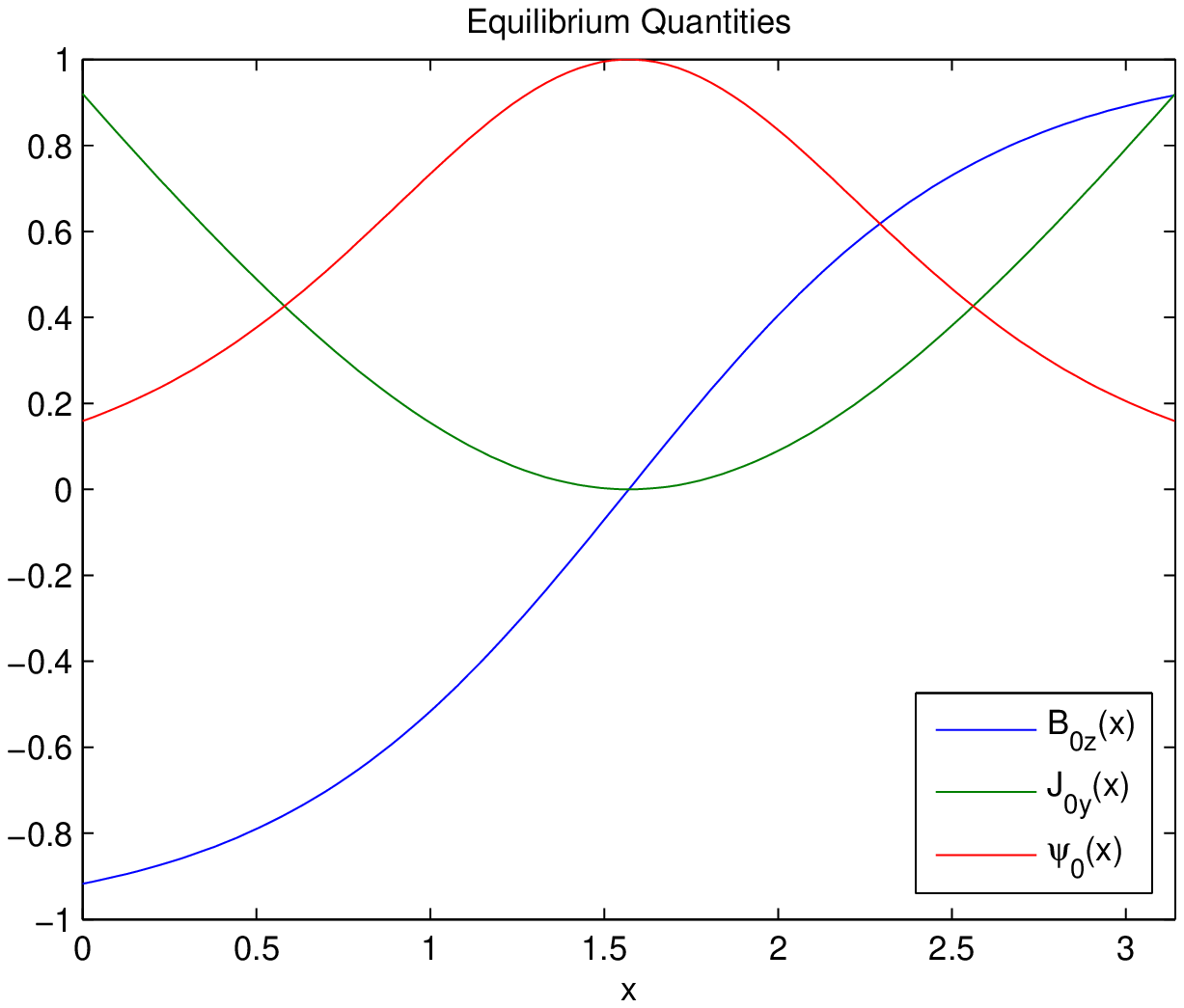}
\caption{The equilibrium profiles of magnetic field, $B_{0z}$, current density, $J_{0y}$ and flux function $\psi_{0}$. \label{fig1}}
\end{figure}

The governing equations defining temporal evolution of perturbations is a set of single-fluid MHD equations:
\begin{equation}
\label{eq:psi}
 \frac{\partial\psi}{\partial t} = [\phi, \psi] + \eta\nabla^{2}\psi,
\end{equation}
\begin{equation}
\label{eq:phi}
 \frac{\partial\nabla^{2}\phi}{\partial t} = [\nabla^{2}\psi, \psi]
\end{equation}
where $\psi$, the magnetic flux and $\phi$, the stream function are defined as follows:
\begin{equation}
\label{eq:perturb1}
 \textbf{B}(x,z,t) = \nabla\psi(x,z,t) \times \hat{j} + b(x,z,t)\hat{j},
\end{equation}
\begin{equation}
\label{eq:pertur}
\textbf{V}(x,z,t) = \nabla\phi(x,z,t) \times \hat{j} + v_{y}(x,z,t)\hat{j}.
\end{equation}
The Poisson bracket notation $[A, B]$=$(\nabla A \times \nabla B)\cdot\hat{k}$ is adopted in Eqs.~\ref{eq:psi} and ~\ref{eq:phi}. Thus,
the linearized dimensionless MHD equations in terms of $\psi$ and $\phi$ are given by:
\begin{eqnarray}
\label{eq:mag}
   \frac{\partial\psi}{\partial t} = \frac{\partial\phi}{\partial z} \frac{\partial\psi_{0}}{\partial x}- \eta \nabla^{2}\psi
   \end{eqnarray}
\begin{eqnarray}
\label{eq:velo}
  \frac{\partial \nabla^{2}\phi}{\partial t} = \frac{\partial \nabla^{2}\psi}{\partial z} \frac{\partial\psi_{0}}{\partial x}-\frac{\partial \psi}{\partial z}\frac{\partial \nabla^{2}\psi_{0}}{\partial x}
\end{eqnarray}
where the partial differences are obtained from Eqs.~\ref{eq:perturb1} and ~\ref{eq:pertur} as follows:
\begin{equation}
\label{eq:per}
\frac{\partial\phi}{\partial z}= -v_{x}, \nonumber\\
\frac{\partial\psi_{0}}{\partial x}= B_{0z}, \nonumber\\
j_{0y}= \nabla^{2}\psi_{0}.
\end{equation}
Here, velocity, magnetic field, magnetic flux, stream function, time and space coordinates are normalized to $V_{A0}$, $B_ {\rm 0}$, $B_{\rm 0} a$, $a^2/\tau_{A}$,
$\tau_{A}$, and $a$ (spicule radius), respectively.
The equilibrium profiles are shown in Figure~\ref{fig1}.

\section{Numerical results and discussion}
We use the finite difference and the Fourth-Order Runge-Kutta methods to take the space and time derivatives in the coupled
Eqs.~\ref{eq:mag} and ~\ref{eq:velo}. The implemented numerical scheme is using by the forward finite difference method
to take the first spatial derivatives with the truncation error of ($\Delta x$), which is the spatial resolution in the $x$ direction.
The order of approximation for the second spatial derivative in the finite difference method is $O((\Delta x)^2)$.
On the other hand, the Fourth-order Runge-Kutta method takes the time derivatives in the questions. The computational output data
are given in $17$ decimal digits of accuracy.
\\We set the number of mesh-grid points as~$256\times256$. In addition, the time step is chosen as $0.001$, and the system length
in the $x$ and $z$ dimensions (simulation box sizes) are set to be ($0$,$\pi$) and ($0$,$2\pi$).

The parameters in spicule environment are as follows:
$a=200$~km(spicule radius), $L=6000$~km (Spicule length), $v_{0}=25$~km s$^{-1}$,
$n_{e}=11.5\times10^{16}$~m$^{-3}$, $B_{0}=1.2\times10^{-3}$~Tesla, $T_{0}=14~000$~K, $g=272$~m s$^{-2}$, $R=8300$~m$^{2}$s$^{-1}$k$^{-1}$
(universal gas constant), $V_{A0}=77$~km/s, $\mu=0.6$, $\tau= 2.5$~s, $\rho_{0}=1.9\times10^{-10}$~kg m$^{-3}$,
$p_{0}=3.7\times10^{-2}$~N m$^{-2}$, $\mu_{0}=4\pi \times10^{-7}$~Tesla m A$^{-1}$, $z_{w}=0.5$ and $z_{t}=1.57$ (in our dimensionless units),
$H= 400$~km, $\eta=10^{3}$~m$^2$ s$^{-1}$, and $k= 2\pi/L_{z}= 1$ (dimensionless wavenumber normalized to $a$).

\begin{figure*}
\centering
\includegraphics[width=17cm, height=15cm]{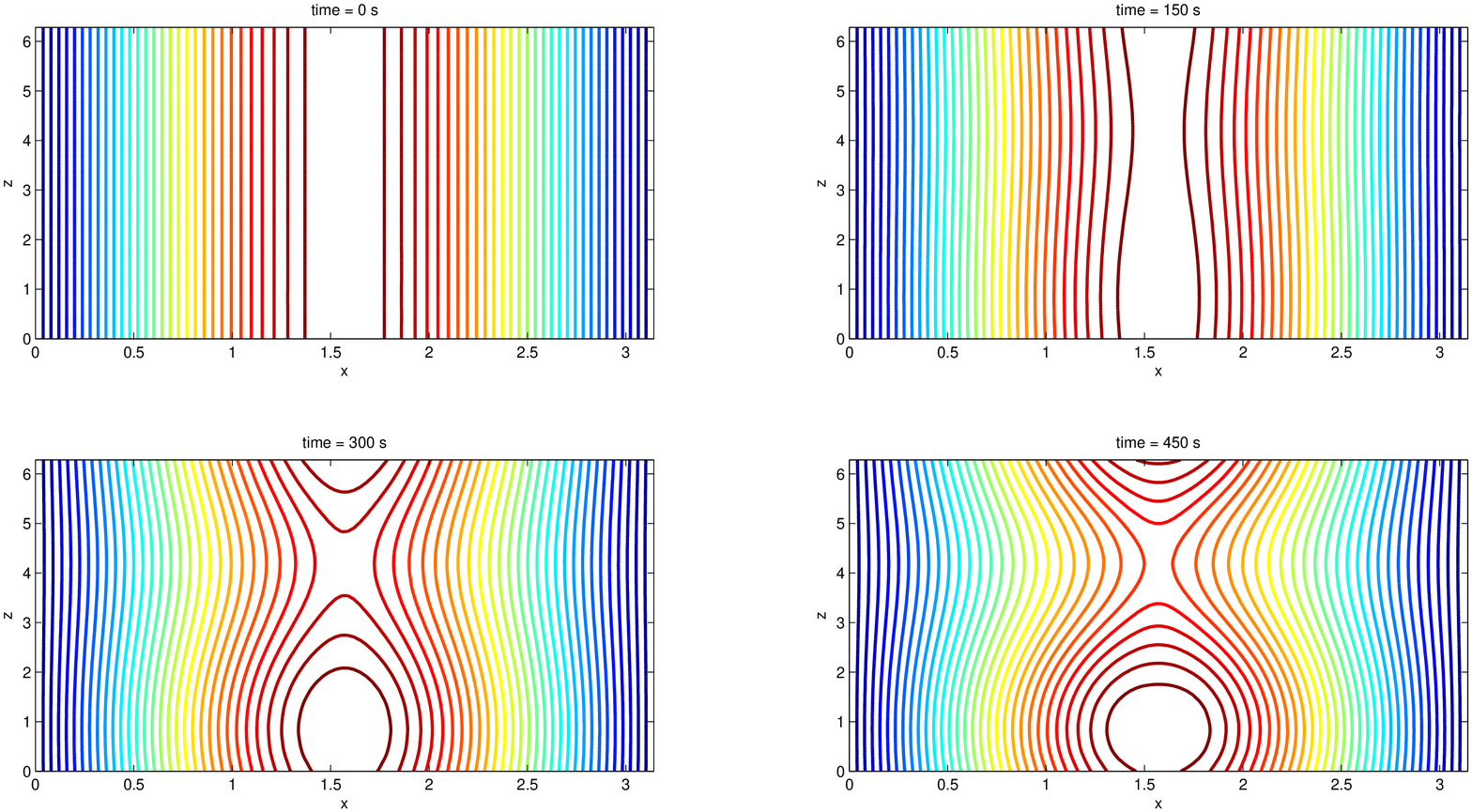}
\caption{Contour plots of perturbed magnetic flux function at different times. \label{fig2}}
\end{figure*}

Figure~\ref{fig2} shows the contour plots of the perturbed flux function at different times, $t$= $0~s$, $t$= $150~s$, $t$= $300~s$ and $t$= $450~s$.
As tearing instability proceeds, magnetic islands continuously emerge and develop as a result of ongoing magnetic field line reconnection in the
considered lifetime of a solar spicule. In this figure, the formation of magnetic islands characterized by the magnetic X and O points can be seen clearly there.
The growth rate of tearing mode instability can be numerically obtained through calculating the slope of energy-time graph at linear stage.

\begin{figure}
\centering
\includegraphics[width=8cm]{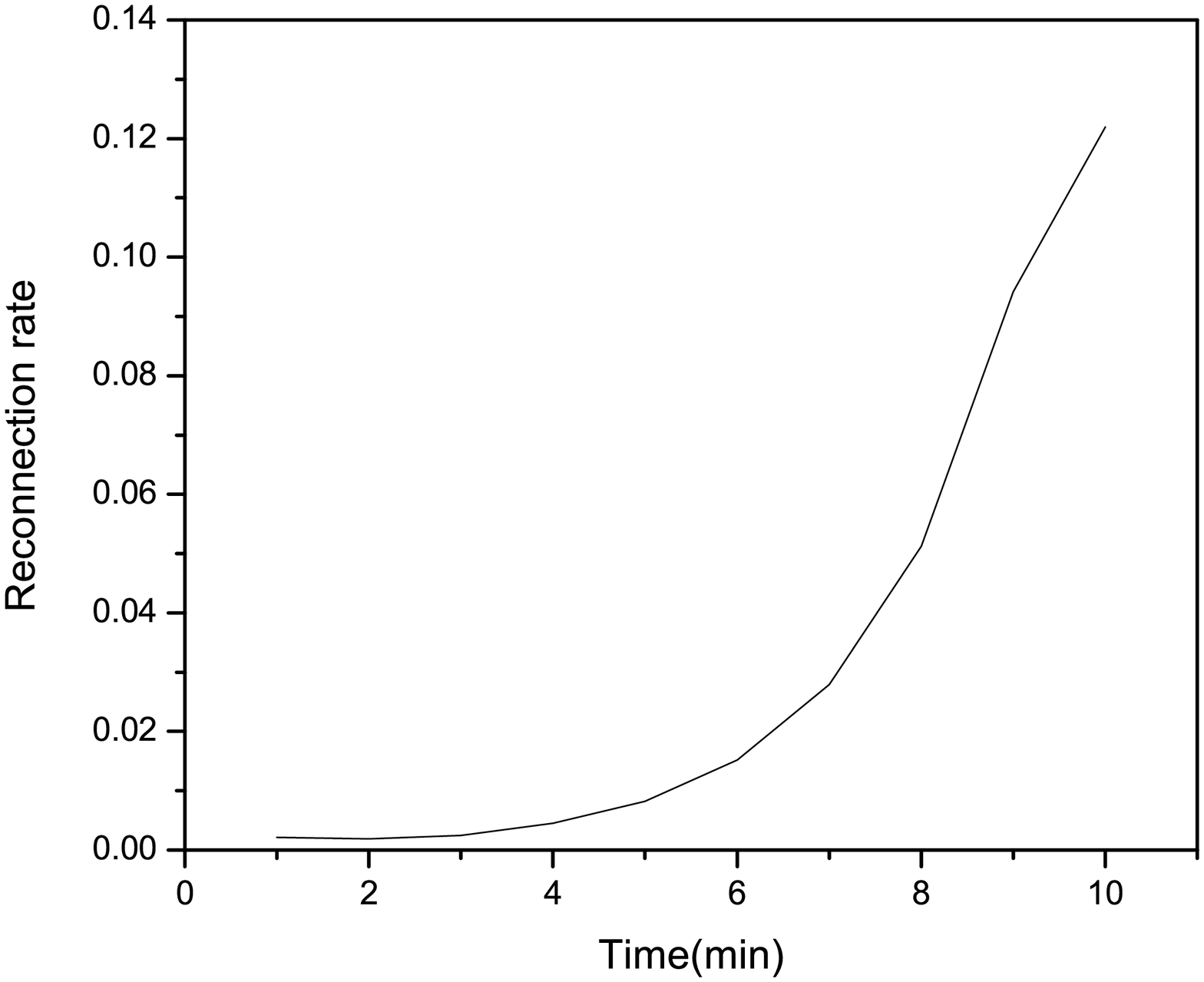}
\caption{Time evolution of the magnetic reconnection rate. \label{fig3}}
\end{figure}

Figure~\ref{fig3} shows temporal variations of the magnetic reconnection rate. The reconnection rate is obtained by certifying the magnetic reconnection site
at each time step, which is found from the value of the magnetic flux at the point $(128, 163)$ (the reconnection cite in the simulation box). In this figure,
the reconnection rate increases exponentially during a spicule life time which is due to the linear phase of the instability \citep{Yoko1996}.
The gas density decreases at neutral point because of gravitational downflow and of plasma rarefaction due to outflows accelerated by the magnetic
tension of the reconnected field lines when island is formed. Our simulations show that the reconnection can be occurred during of a spicule life time.
It is demonstrated by \citet{He2009} observationally in a typical limb spicule.

\begin{figure}
\centering
\includegraphics[width=8cm]{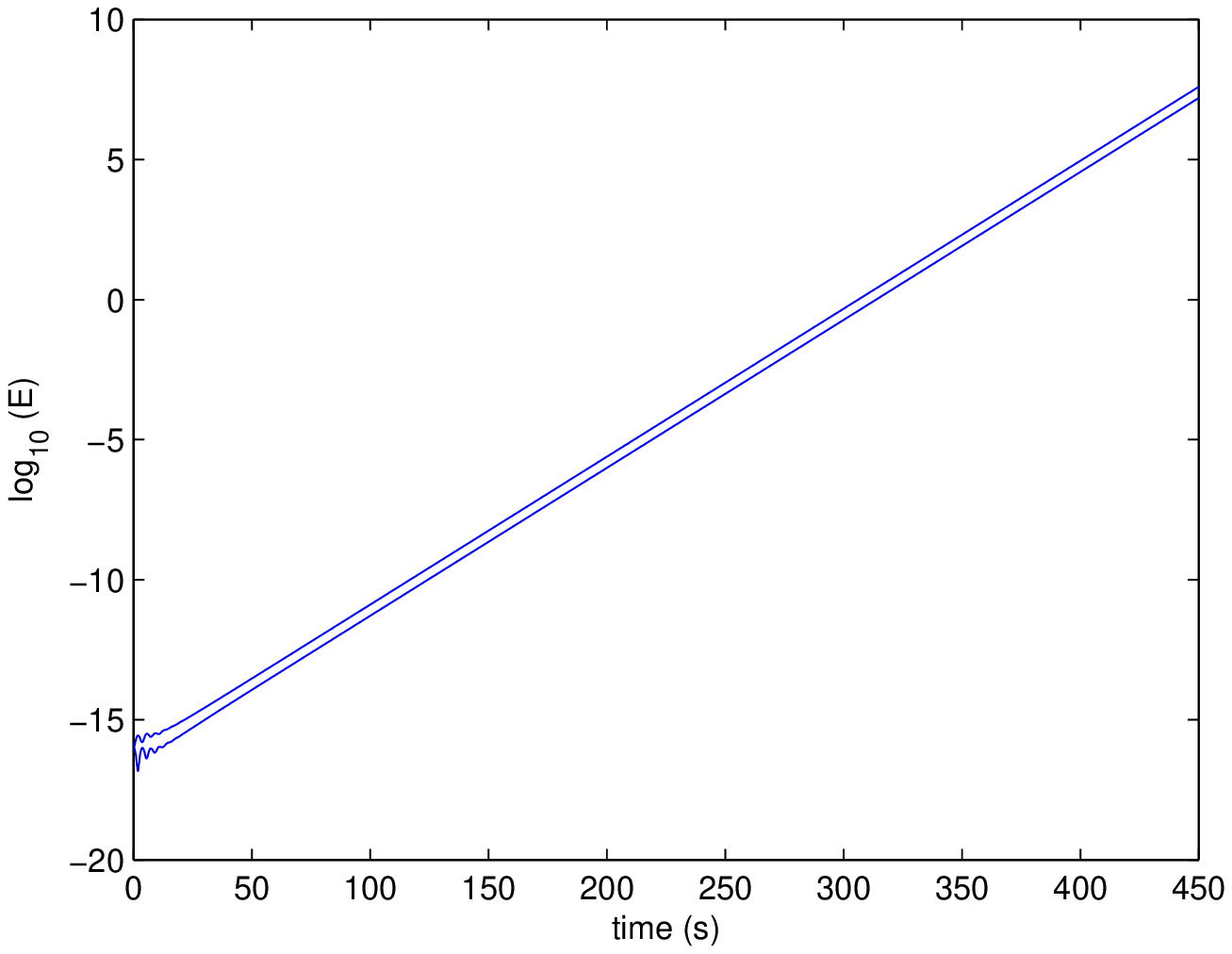}
\caption{Time evolution of magnetic and kinetic energies in the linear phase of the reconnection. \label{fig4}}
\end{figure}

The logarithmic energy-time graph which has been shown in Figure~\ref{fig4}, display the time variation of dimensionless magnetic
energy, $E_{m}$=$\int(B_{x}^{2} + B_{z}^{2})dv$ and kinetic energy, $E_{k}$=$\int(v_{x}^{2} + v_{z}^{2})dv$. In the linear phase, the
logarithmic magnetic and kinetic energies grow linearly (or exponentially in the non-logarithmic case) with time.

\section{Conclusion}
\label{sec:concl}
The linear dynamics of magnetic reconnection in the spicule structure was studied numerically.
A set of incompressible magnetohydrodynamic (MHD) equations described the evolution of tearing instability
of a slab of uniform plasma which had been embedded in a force-free equilibrium magnetic field.
The resistivity of plasma was the main agent to trigger the reconnection process by breaking
of the frozen-in field constraint. The considered plasma medium in this study was the solar
spicule environment.
In fact, the ongoing reconnection process is intrinsically associated with the formation of
magnetic islands, and in the linear phase of instability, the magnetic and kinetic energies
in logarithmic scale grow linearly as it is expected.

\acknowledgments
This work has been supported financially by the Research Institute for Astronomy and
Astrophysics of Maragha (RIAAM), Maragha, Iran.

\makeatletter
\let\clear@thebibliography@page=\relax
\makeatother

\end{document}